\documentclass[12pt]{iopart}
\usepackage{iopams}
\usepackage{graphicx}

\begin{document}

\title[Electronic structure and transport properties of YPtIn and LuAgGe]{%
Electronic structure and anisotropic transport properties in hexagonal YPtIn
and LuAgGe ternary compounds}

\author{G. D. Samolyuk, S. L. Bud'ko, E. Morosan, V. P. Antropov and P. C.
Canfield}

\address{Ames Laboratory US DOE and Department of Physics and Astronomy,
Iowa State University, Ames, IA 50011, USA}

\begin{abstract}
We present anisotropic, zero applied magnetic field, temperature dependent
resistivity measurements on hexagonal, non-magnetic, YPtIn and LuAgGe single
crystals. For these materials the in-plane resistivity, $\rho_{ab}$, is
significantly higher than the $c$ - axis one, $\rho_c$, with $%
\rho_{ab}/\rho_c \approx 1.4$ for YPtIn and $\approx 4.2 - 4.7$ for LuAgGe. The connection between the electronic
structure and the anisotropic transport properties is discussed using density functional calculations that link
the observed anisotropy with a specific shape of Fermi surface and anisotropy of the Fermi velocities.
\end{abstract}


\submitto{\JPCM}

\maketitle

\section{Introduction}

Interest to the RAgGe and RPtIn (R = rare earth) ternary intermetallics \cite%
{gib96a,fer74a} was recently stimulated by growth of single crystals of these materials and in-depth
characterization of their thermodynamic and transport properties \cite{mor04a,mor05a}. Due to the curious
combination of crystallographic (hexagonal) and point (orthorhombic) symmetries of the rare earths in these two
families, local magnetic moment members of the series have complex (but tractable) anisotropic $H - T$ phase
diagrams, with intriguing metamagnetism as well as crystal electric field effects playing significant role in
defining of the magnetic order \cite{mor05a,mor05b}. YbAgGe and
YbPtIn were described as heavy fermion compounds \cite%
{mor04a,kat04a,tro00a,kac00a} with YbAgGe being a rare example of the Yb -
based stoichiometric compound demonstrating field-induced non-Fermi-liquid
behavior \cite{bud04a,bud05a}. YbPtIn displays similar properties but with a
less comprehensively determined phase diagram \cite{mor05c}.

The non-magnetic members of these families, LuAgGe and YPtIn, have rather unpretentious physical properties
\cite{mor04a,mor05a}, however, understanding of these materials from the band-structure point of view can be
viewed as the first step towards comprehension of the rest of the series, especially since they serve as
non-magnetic references in the analysis of the physical properties across the series. In particular, the
anisotropic crystal structure suggests measurable anisotropy of the transport properties. Comparison of the
experimental resistivity data with computational, band-structure results will serve as a test of the theoretical
description of these materials and will allow for the identification of the qualitative features of their Fermi
surfaces responsible for the anisotropy.

\section{Experimental methods and computational details}

LuAgGe and YPtIn single crystals in the form of hexagonal-cross-section rods of several mm length and up to 1
mm$^2$ cross section were grown from high temperature ternary solutions (rich in Ag and Ge for LuAgGe and in In
for YPtIn), their structure and the absence of impurity phases were confirmed by powder X-ray diffraction (see
\cite{mor04a,mor05c} for details of the samples' growth and characterization). For resistivity measurements with
$I \| c$, clean, regular-shaped, as-grown or lightly polished rods were used. Samples for $I \| ab$ measurements
were cut out of thicker rods, with their length approximately along [110] crystallographic direction. In the case
of YPtIn particular care was taken to eliminate residues of highly conductive In flux (by polishing with
subsequent etching in HCl) that otherwise could effect the current flowing through the sample (\textit{i.e.}
practically shorting out the sample) resulting in erroneous values for resistivity and its temperature dependence.
The absence of any anomaly in resistivity near $T_c^{In} = 3.4$ K served as an indication of a clean sample.

Resistivity measurements were performed over the 2 - 300 K temperature range, in standard four-probe configuration
with thin platinum wires attached to the samples by Epotek H20E silver epoxy, using the ACT option of the Quantum
Design PPMS instrument ($f = 16$ Hz, $I = 1-3$ mA). For each direction of the current several samples were used
and the results were within the error bars ($\sim 20\%$) of the measurements of the dimensions of the samples and
the position of the voltage leads.

\begin{figure}[tbp]
\begin{center}
\includegraphics[angle=0,width=120mm]{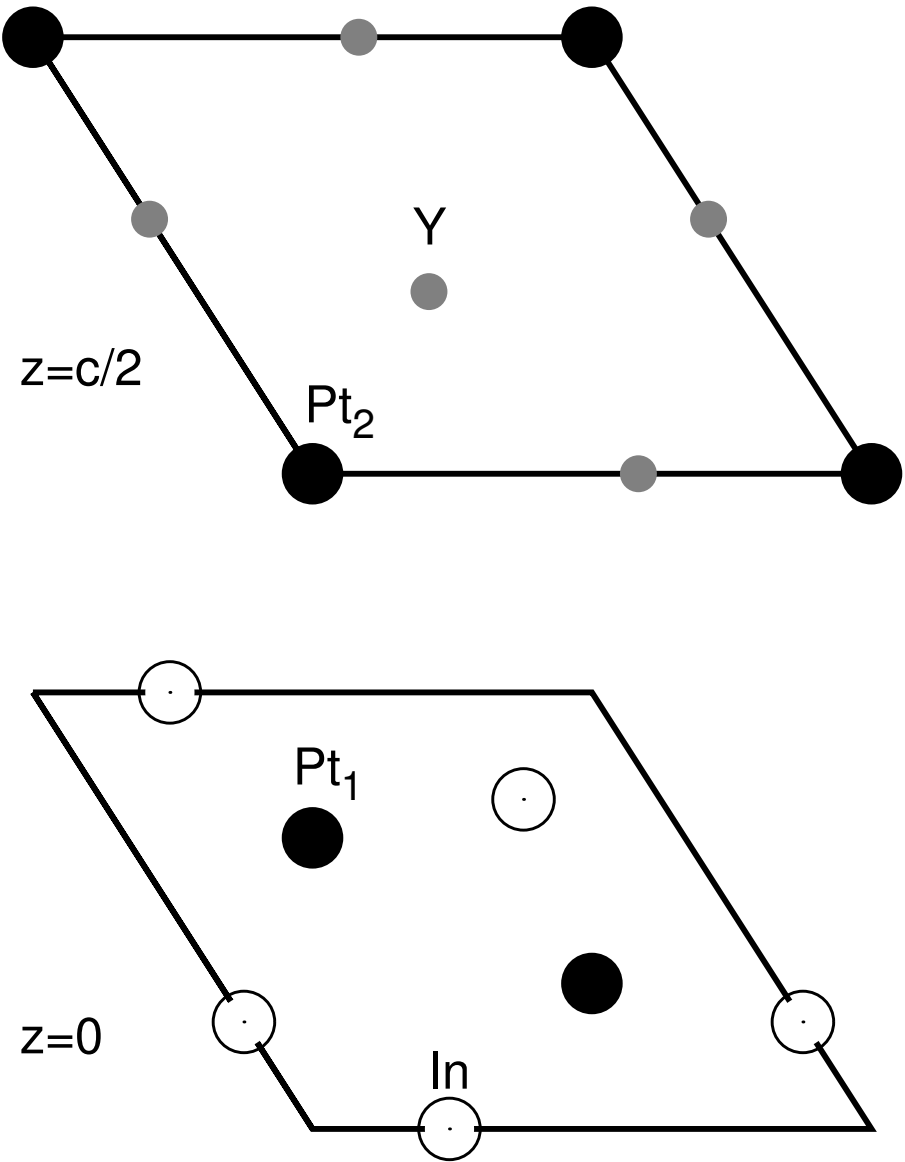}
\end{center}
\caption{The unit cell of YPtIn (hexagonal structure of ZrNiAl-type).}
\label{fig:struct}
\end{figure}

Both, LuAgGe \cite{gib96a} and YPtIn \cite{fer74a} crystallize in Fe$_2$%
P/ZrNiAl-type of structure (\textit{P62m} space group) with nine atoms per
unit cell (Fig. \ref{fig:struct}). The $z=0$ plane is occupied by Pt(Ge),
In(Ag) atoms, whereas $z=c/2$ plane by Y(Lu), Pt(Ge) atoms. The apparent
layered character of this structure may lead to strong anisotropy of Fermi
surface.

Electronic structure was calculated using the atomic sphere approximation tight binding linear muffin-tin orbital
(TB-LMTO-ASA) method \cite{and75a,and84a} within the local density approximation (LDA) with Barth-Hedin
\cite{bar72a} exchange-correlation at experimental values of the lattice parameters. A mesh of 4699 $\vec{k}$
points in the irreducible part of the Brillouin zone (BZ) was sufficient to reach a few percent accuracy in the
calculated conductivity tensor. $3d$ electrons of Ge and $4d$ electrons of In were included in the core states.
$4f$ electrons of Lu atom were treated as the valence states.

With the purpose of investigating the anisotropic behavior of transport
properties, we implement the expression for diffusion conductivity in the
relaxation time approximation \cite{zim67a}
\begin{equation}  \label{eq:one}
\sigma_{\alpha}(E) \propto \tau\sum\limits^{}_{\vec
k,\nu}v^{\alpha}_{\nu}(\vec k)v^{\alpha}_{\nu}(\vec k)\delta[%
\varepsilon_{\nu}(\vec k)-E] \equiv \tau<v^{\alpha}v^{\alpha}>
\end{equation}
In Eq. \ref{eq:one}, $\tau$ is the relaxation time, $v^{\alpha}_{\nu}=%
\partial \varepsilon_{\nu}(\vec k)/\partial \vec k$ is the electronic group
velocity, $\varepsilon_{\nu}(\vec k)$ is the energy spectrum, $\vec k$ and $%
\nu$ are the wave vector and band index, and $E$ corresponds to Fermi energy ($E_F$).

\section{Results and discussion}

\subsection{Experiment}

\begin{figure}[tbp]
\begin{center}
\includegraphics[angle=0,width=100mm]{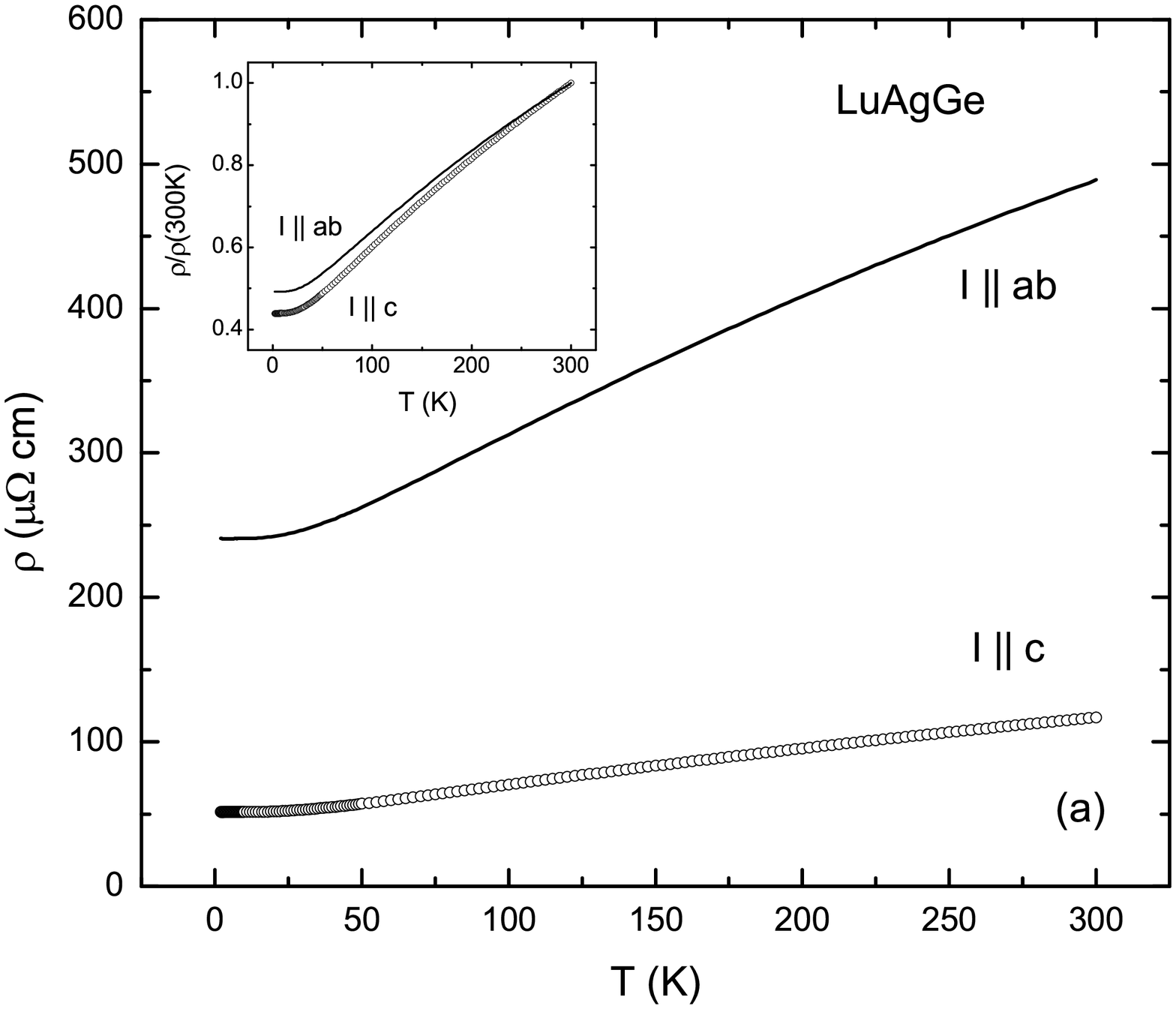} %
\includegraphics[angle=0,width=100mm]{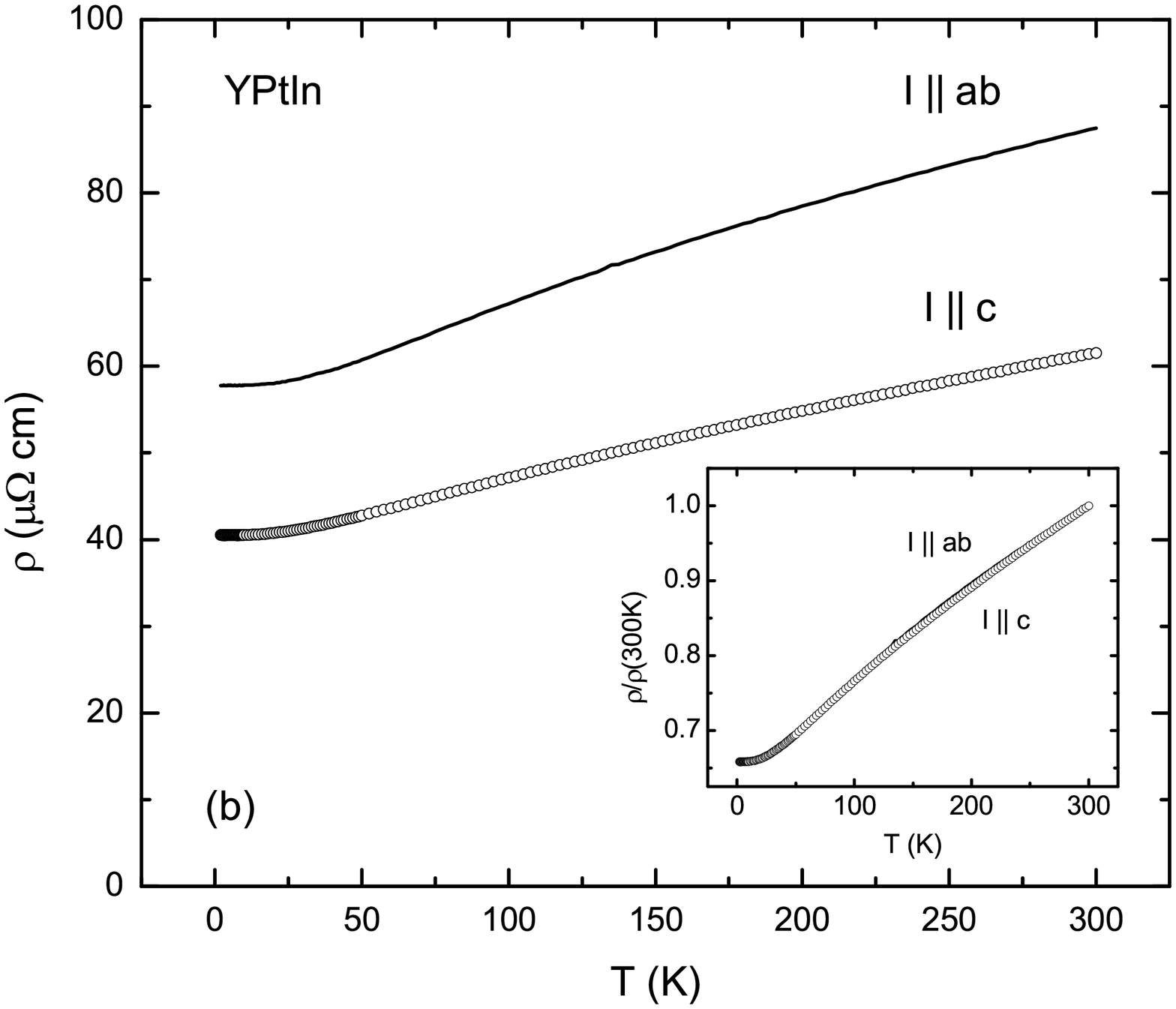}
\end{center}
\caption{Anisotropic zero field resistivity of (a) LuAgGe, (b) YPtIn.
Insets: the same data normalized to $\protect\rho(300$K$)$.}
\label{rho}
\end{figure}

Zero field resistivity data for different directions of the current flow in LuAgGe and YPtIn are shown in Fig.
\ref{rho}. For both materials $\rho_{ab} > \rho_c$ although in both cases the normalized resistivities (see insets
to Fig. \ref{rho} (a),(b)) are practically isotropic. The measured anisotropy in resistivity at 300 K is 4.2
(LuAgGe) and 1.4 (YPtIn). It should be noted that given the aforementioned (conservative) estimate of the error
bars for the absolute values of resistivity, the $\rho(T)$ for YPtIn may be close to being isotropic, whereas for
LuAgGe the inequality $\rho_{ab} > \rho_c$ is unambiguous. The residual resistivity ratios (RRR) are quite low,
$1.5 - 2$, similarly mediocre RRR were observed for the other members of these families \cite{mor04a,mor05a}.
Anisotropic susceptibility and electronic coefficient
of the specific heat of these two materials were reported earlier \cite%
{mor04a,mor05a}. For LuAgGe $\chi_{ab}(300K) \approx -3.3~10^{-5}$ emu/mol, $%
\chi_{c}(300K) \approx -5~10^{-5}$ emu/mol, $\gamma \approx 1.4$ mJ/mol K$^2$%
; for YPtIn $\chi_{ab}(300K) \approx 6.6~10^{-5}$ emu/mol, $\chi_{c}(300K)
\approx 4.6~10^{-5}$ emu/mol, $\gamma \approx 6.7$ mJ/mol K$^2$. The
magnetic susceptibility for both compounds is slightly anisotropic and on a
gross level temperature-independent (except for the low temperature, small,
impurity tail). Since it is quite difficult to quantitatively account for
different contributions to susceptibility (Landau, core, \textit{etc.}) (see
\textit{e.g.} \cite{sch97a}) we will not attempt to compare the measured
values with the band structure calculations. The electronic coefficient of
the specific heat, $\gamma$, on the other hand, can be useful for evaluating
the band structure calculations.

\subsection{Electronic structure}

\begin{figure}[tbp]
\begin{center}
\includegraphics[angle=0,width=120mm]{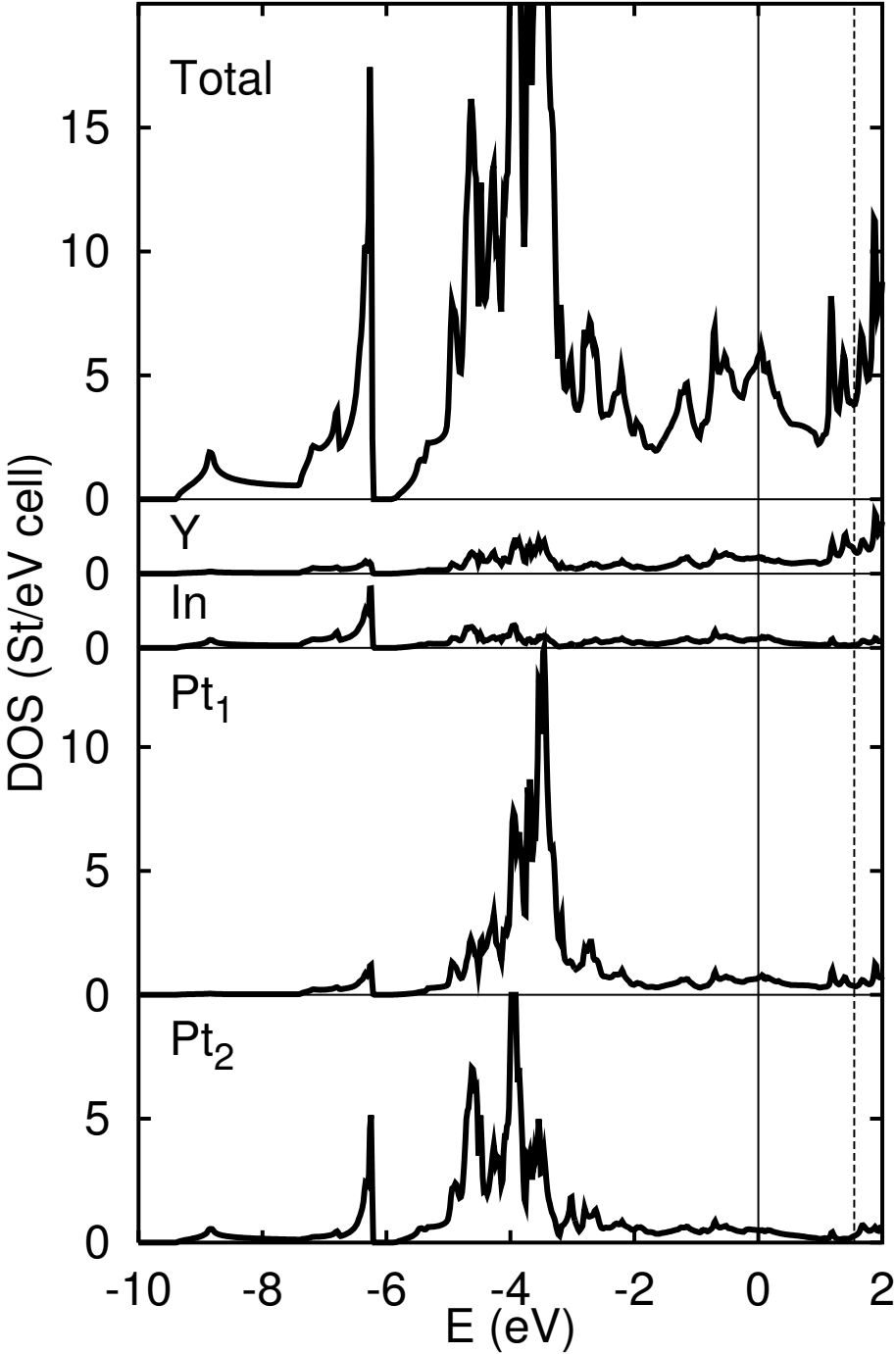}
\end{center}
\caption{ The total and partial DOS for the YPtIn. $E_F$ corresponds to zero
energy (the unit cell contains 48 valence electrons). The energy, which
corresponds to 54 valence electrons in unit cell (number of valence
electrons for LuAgGe without Lu $f$-electrons) is shown by the dashed line.
The vertical scales are the same for all panels.}
\label{fig:dos_yptin}
\end{figure}

\begin{figure}[tbp]
\begin{center}
\includegraphics[angle=270,width=120mm]{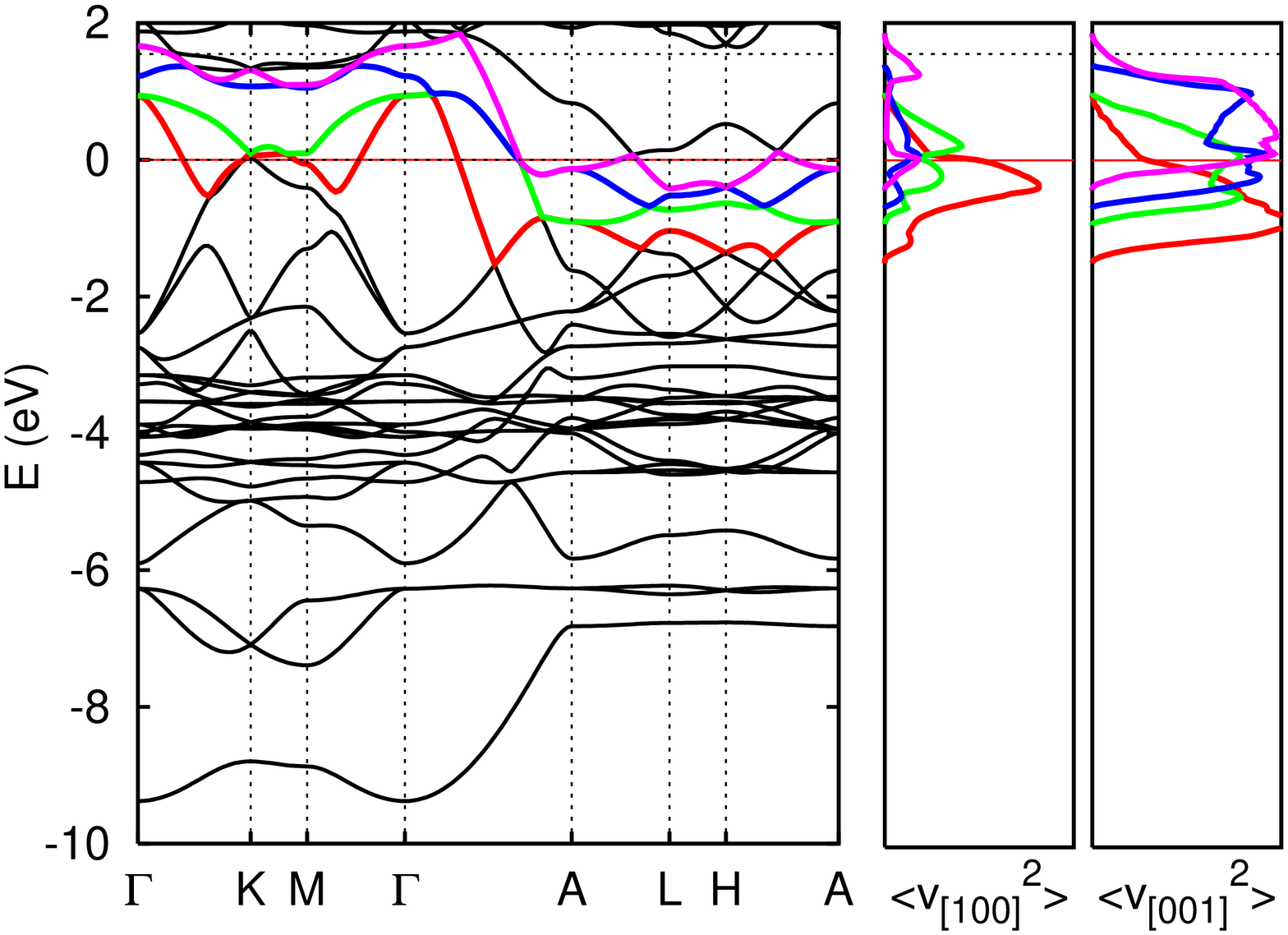}
\end{center}
\caption{ The band structure (left panel) and [100], [001] components of
velocity tensor as a function of energy (right panels) for YPtIn. $E_F$
corresponds to zero energy (the unit cell contains 48 valence electrons).
The $E_F$, which corresponds to 54 valence electrons is shown by horizontal
dashed line.}
\label{fig:bnds_yptin}
\end{figure}

The calculated total and partial densities of states (DOS) for YPtIn is
shown in Fig. \ref{fig:dos_yptin}. The sharp peak at -6.5 eV below the Fermi
level ($E_F$) corresponds to In-$s$ bands (Fig. \ref{fig:bnds_yptin}). These
states are strongly hybridized with $s$-states of Pt atoms. A gap of 0.4 eV
separates the In- and Pt-$s$ states from the conductivity electrons band. The Pt-$%
d$ states are localized at -4 eV and are hybridized with $d$-states of Y and
$p$-states of In atoms. At $E_F$ the contribution of the electrons
originating from all atoms is nearly the same and is proportional to the
number of the corresponding atoms in the unit cell. The electronic specific
heat coefficient $\gamma_{band}$ calculated from DOS at $E_F$ ($N(E_F)$)
(Table~\ref{tab:results}) is 4.2 mJ/mol K$^2$.

\begin{figure}[tbp]
\begin{center}
\includegraphics[angle=0,width=120mm]{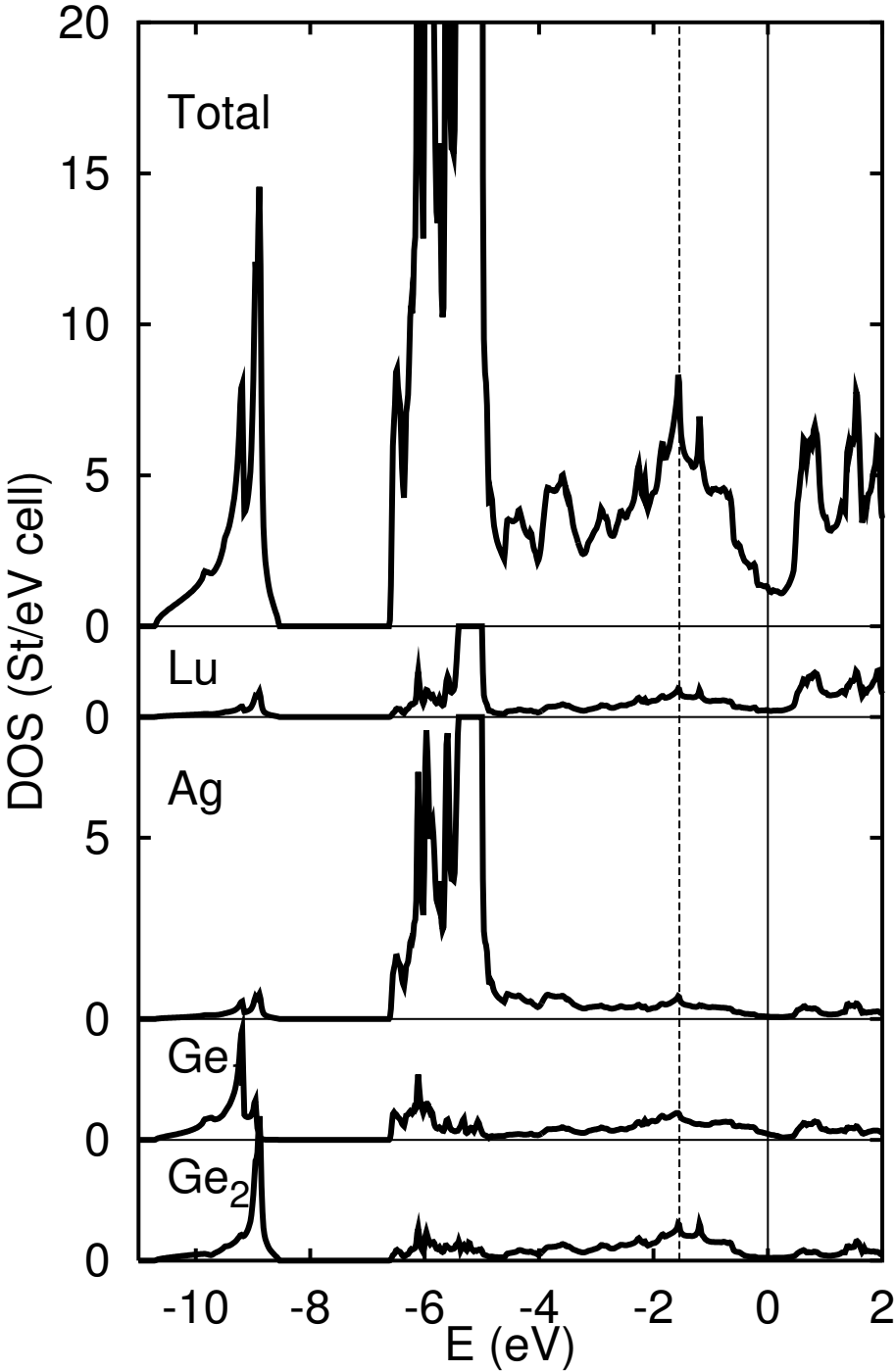}
\end{center}
\caption{ The total and partial DOS for the compound LuAgGe. $E_F$
corresponds to zero energy (the unit cell contains 96 valence electrons).
The $E_F$, which corresponds to 90 valence electrons is shown by vertical
dashed line.}
\label{fig:dos_luagge}
\end{figure}

In the LuAgGe compound the Ge$-s$ states band is separated from conductivity
electrons band by 2 eV gap (Figs. \ref{fig:dos_luagge} and \ref%
{fig:bnds_luagge}). This value is significantly larger compared to YPtIn. The difference in the gap size is due to
the deeper position of the Ge-$s$
electrons states than In and smaller width of $d$-band in 4$d$- compared to 5%
$d$-metals~\cite{and70a}. The Lu-4$f$ states are located around 5 eV below $%
E_F$ within the range of the Ag-$d$ states and they show a very little
dispersion. In contrast to YPtIn the $E_F$ in LuAgGe compound is placed in
pseudo-gap. The $N(E_F)$ equals to 1.92 St/(Ry atom). This decrease of $%
N(E_F)$ translates into four times smaller $\gamma_{band}$ value compared to
YPtIn (Table~\ref{tab:results}). The main contribution to $N(E_F)$ is coming
from the $d$-states of Lu and $p$-states of Ge1 atoms located in the same
plane as Lu.

The comparison of the calculated, $\gamma_{band}$, and measured, $%
\gamma_{exp}$, electronic specific heat coefficients (Table \ref{tab:results}%
) as $\gamma_{exp} = \gamma_{band} (1 + \lambda)$ give the value of the
enhancement factor $\lambda$ (due to the electron-electron and
electron-phonon interactions) of 0.4-0.6. Such $\lambda$ values are very
common for ordinary metals and were observed \textit{e.g.} for several
alkali metals \cite{gri81a}, being attributed mainly to the electron-phonon
interactions.

To demonstrate the similarity of the electronic structures for YPtIn and
LuAgGe we calculated the $E_F$ position using YPtIn DOS for 54 valence
electrons per unit cell (this number of electrons corresponds to LuAgGe
compound without Lu-$4f$ states) (see vertical dashed line on Fig.~\ref%
{fig:dos_yptin}). The obtained $E_F$ equals to 1.8 eV and is placed very close to the pseudo-gap in the DOS. The
$E_F$ calculated from LuAgGe DOS (Fig.~\ref{fig:dos_luagge}) for 90 electrons per cell (this number of electrons
corresponds to hypothetic compound with 48 electrons from YPtIn ligands plus additional $3 \times 14 = 42$
$f$-electrons from the rare earth) equals to -1.8 eV and is placed at the peak in DOS very similar to the one in
YPtIn. These results suggest that the rigid-band approximation could be a reasonable approach for the analysis of
the conductivity as a function of the band filling.

\begin{figure}[tbp]
\begin{center}
\includegraphics[angle=270,width=100mm]{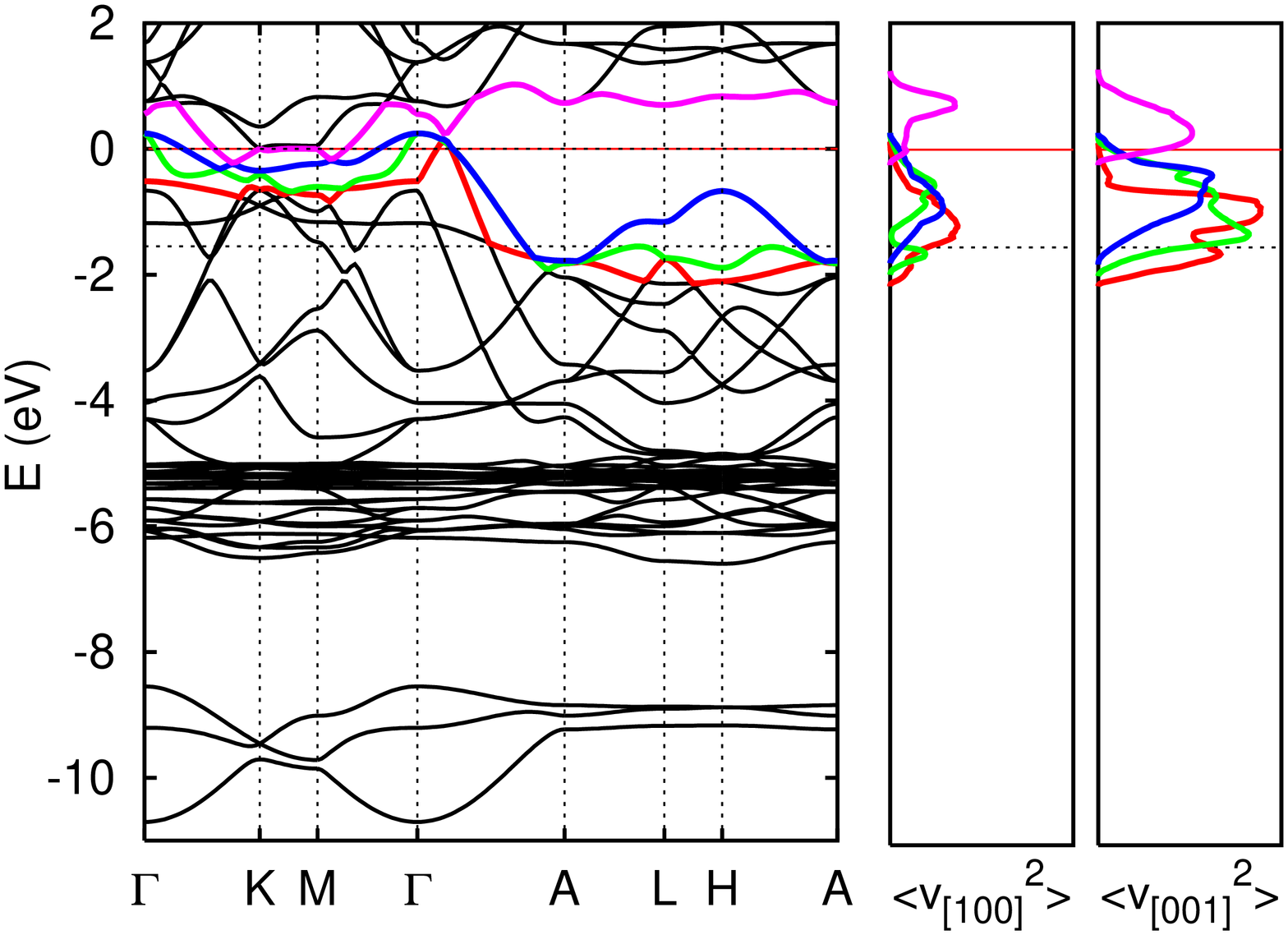}
\end{center}
\caption{ The band structure (left panel) and [100], [001] components of
velocity tensor as a function of energy (right panels) for LuAgGe. $E_F$
corresponds to zero energy (the unit cell contains 96 valence electrons).
The $E_F$, which corresponds to 90 valence electrons is shown by horizontal
dashed line.}
\label{fig:bnds_luagge}
\end{figure}

The partial contributions from the different bands to the velocity tensor $%
<v^{\alpha}v^{\alpha}>$ for in-plane ([100]-direction) and perpendicular to
plane ([001]-direction) are shown in the left panels of Figs.~\ref%
{fig:bnds_yptin} and \ref{fig:bnds_luagge}. The sizable difference between
velocity tensor in [001] and [100] directions is caused by larger dispersion
of bands crossing Fermi level in $\Gamma$-$A$ direction compared to $\Gamma$-%
$K$. Whereas it is difficult to single out the band responsible for the conductivity anisotropy in YPtIn, for
LuAgGe it is the Lu-$d$, Ge1-$p$ band shown in purple in Fig.~\ref{fig:bnds_luagge}.

\begin{figure}[tbp]
\begin{center}
\includegraphics[angle=0,width=110mm]{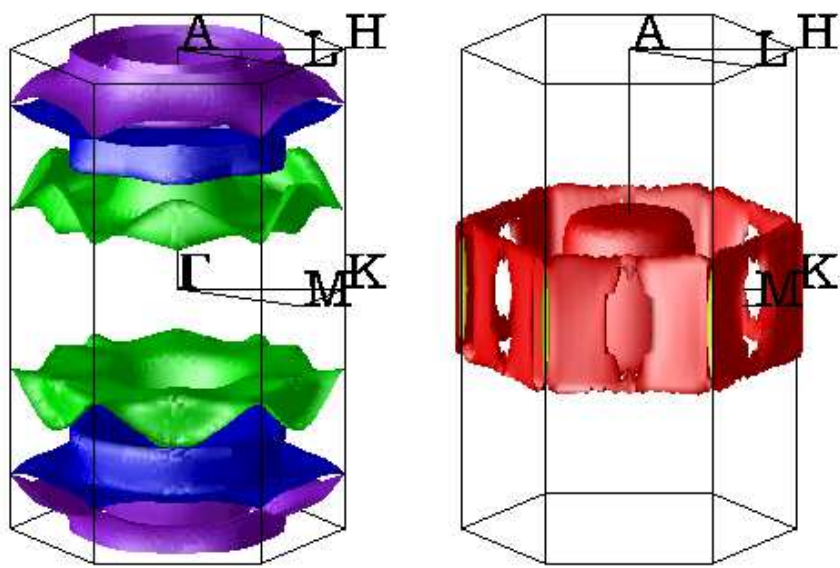}
\end{center}
\caption{ The Fermi surface of YPtIn. The colors of the different sheets correspond to the bands colors in
Fig.~\ref{fig:bnds_yptin}.} \label{fig:fs_yptin}
\end{figure}

This anisotropy of the Fermi velocity distribution is caused by a mainly two-dimensional shape of the Fermi
surface (FS) of YPtIn (the Fermi velocity is perpendicular to the Fermi surface) (Fig.~\ref{fig:fs_yptin}). At
least two of the four FS sheets are open quasi-two-dimensional surfaces perpendicular to the [001]-direction. Such
a shape of the FS sheets is in turn determined by the layered character of YPtIn, where layers of Y-Pt2 atoms are
separated by layers of In-Pt2 atoms.

\begin{figure}[tbp]
\begin{center}
\includegraphics[angle=0,width=110mm]{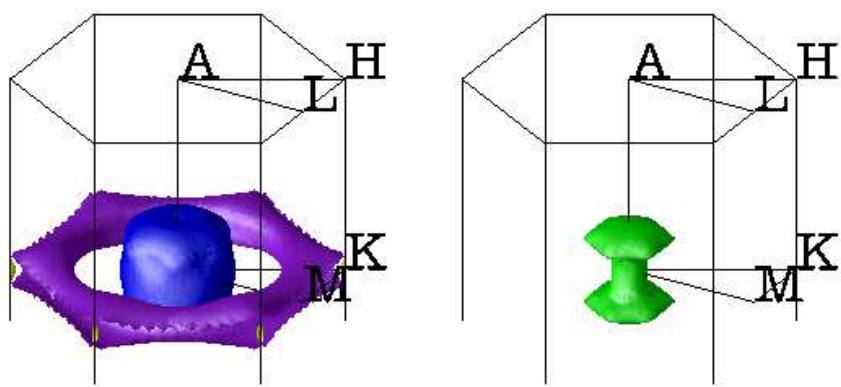}
\end{center}
\caption{ The Fermi surface of LuAgGe. The colors of the different sheets correspond to the bands colors in
Fig.~\ref{fig:bnds_luagge}. Two red strawberry-shaped FS sheets centered near the centers of the dumbbell's disks
are not shown.} \label{fig:fs_luagge}
\end{figure}

The FS of LuAgGe (Fig.~\ref{fig:fs_luagge}) is somewhat different, however
the open quasi-two-dimensional character of one FS sheet is present for this
material as well. This FS sheet is shown in purple in Fig.~\ref%
{fig:fs_luagge} and, as seen from Fig.~\ref{fig:bnds_luagge} its
contribution determines the anisotropic character of conductivity in LuAgGe.

The topology of the Fermi surfaces calculated for YPtIn and LuAgGe (Figs. %
\ref{fig:fs_yptin},\ref{fig:fs_luagge} suggests that search for quantum oscillations in these materials may be
rather complicated. Besides tiny needles centered at $K$-point (in both materials) whose existence may be
questionably because of numerical uncertainties, there are the $\Gamma$-centered dumbbell in LuAgGe and the
$\Gamma$-centered hassock- (marshmallow-) shaped pockets in both materials which may serve as the possible
candidates for de Haas - van Alphen/Shubnikov - de Haas studies. In addition, these pockets (as well as the
quasi-2D sheets in YPtIn) might be studied separately if FS nesting features are analyzed.

The dependency of the [100] and [001] components of the $<v^{\alpha}v^{%
\alpha}>$-tensor as a function of band filling is shown in Fig.~\ref%
{fig:conduct}. The zero energy on these figures corresponds to occupation of 48 electrons calculated from YPtIn
DOS (bottom panel) and 48 plus 3 sets of 14-$f$ electrons for LuAgGe (top panel). The variation of the velocity
tensors as a function of population in these both compounds is very similar.The mentioned applicability of the
rigid band approximation allows us to predict (at least qualitatively) the increase (decrease) of conductivity
anisotropy in YPtIn with substitution of Pt or In atoms by atoms with larger (smaller) number of valence
electrons. The resistivity anisotropy estimated from the calculated $<v^{\alpha}v^{\alpha}>$%
-tensor is in reasonable agreement with experiment: $\rho_{ab} > \rho_c$
with LuAgGe being somewhat more anisotropic than YPtIn (Table~\ref%
{tab:results}).

\section{Summary}

The electronic structure and transport properties were calculated and measured in YPtIn and LuAgGe compounds. The
calculated electronic specific heat coefficient and conductivity anisotropy are in a semi-quantitative agreement
with the experiment. It was shown that the observed anisotropy is a consequence of the shape of the Fermi surfaces
of these compounds. In addition, the applicability of the rigid band approximation to describe the dependence of
the conductivity anisotropy on non-isoelectronic doping is established. The calculated Fermi surfaces of YPtIn and
LuAgGe also can be used for future photoemission and quantum oscillations measurements and other properties of
RPtIn and RAgGe (R = rare earth) series. Finally, we can infer from this work that the Fermi surface topology of
the neighboring (related) members of the RPtIn and RAgGe series will manifest similar complexity.

\ack

Ames Laboratory is operated for the U. S. Department of Energy by Iowa State
University under Contract No. W-7405-Eng.-82. This work was supported by the
director for Energy Research, Office of Basic Energy Sciences. One of the
authors (GDS) would like to thank V.G. Kogan for enlightening discussions.

\clearpage

\begin{table}[tbp]
\caption{Experimental lattice constants in $\mathring{A}$, DOS in states/(Ry
atom), electronic specific heat coefficient in mJ/(mol K$^2$) and
conductivity anisotropy.}
\label{tab:results}%
\begin{indented}
\item[]\begin{tabular}{lccccccc}

\br
Compound&$a$&$c$&DOS&\multicolumn{2}{c}{$\gamma$}&\multicolumn{2}{c}{$\frac{<v^2_{[001]}>}{<v^2_{[100]}>}$}\\
\cline{5-8}
        &   &   &   &calc& exp & calc & exp\\
\br
YPtIn   &7.583&3.846&8.17&4.2&6.7&2.7&1.4\\
\mr
LuAgGe  &7.027&4.127&1.92&1.0&1.4&3.2&4.2\\
\br

\end{tabular}
\end{indented}
\end{table}

\clearpage

\begin{figure}[tbp]
\includegraphics[angle=270,width=120mm]{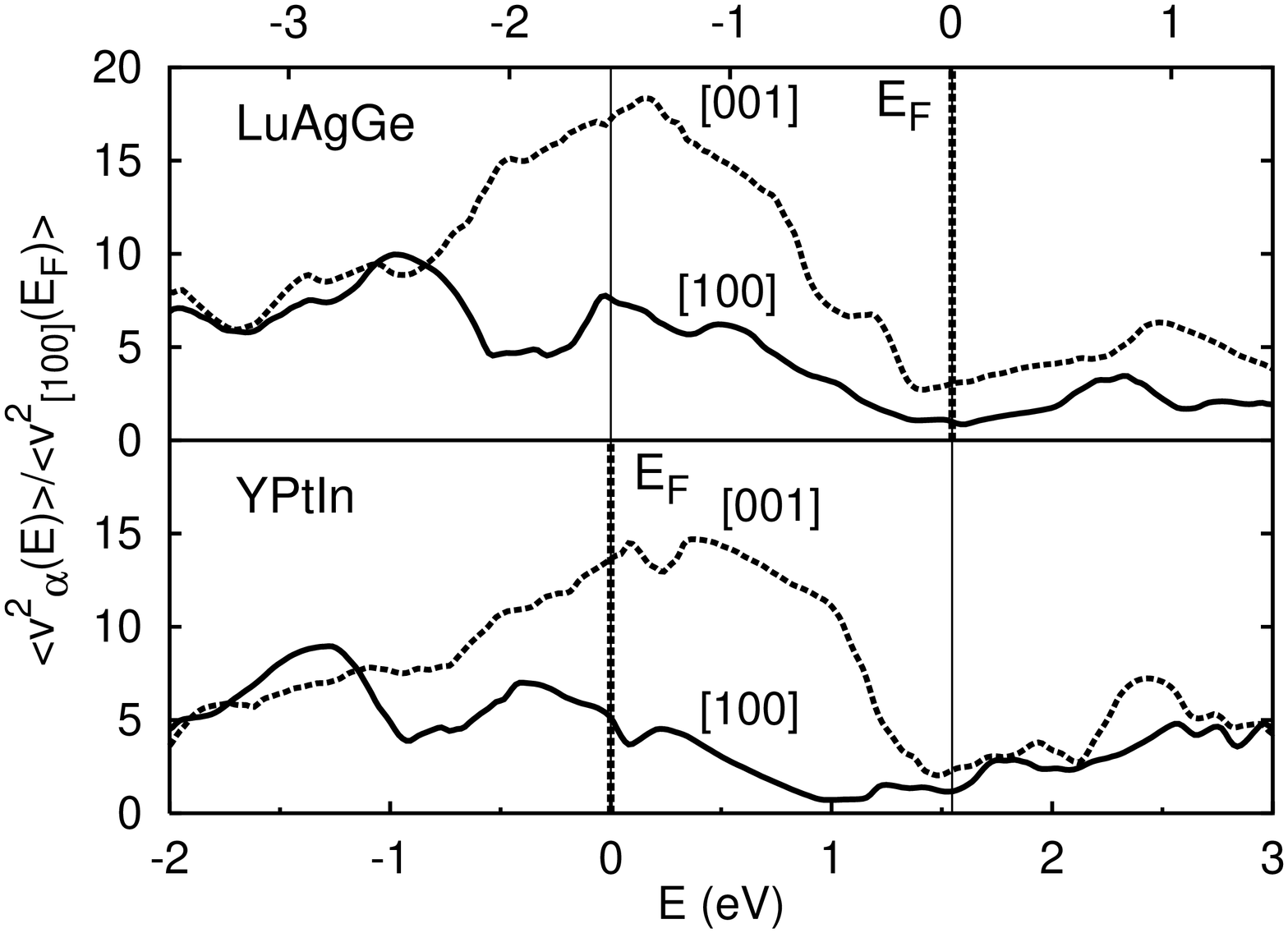}
\caption{ [100], [001] components of the Fermi-velocities tensor as a
function of $E_F$ in YPtIn and LuAgGe compounds. $E_F$ is shown by vertical
dashed line. The position of zero energy is explained in the text.}
\label{fig:conduct}
\end{figure}


\section*{References}

\begin{thebibliography}{99}
\bibitem{gib96a} Gibson B, Poettgen R, Kremer R K, Simon A, Ziebeck K R A
1996 \textit{J. Alloys Comp.} \textbf{239} 34.

\bibitem{fer74a} Ferro R, Marazza R, Rambaldi G 1974 \textit{Zeit. Anorg.
Allgem. Chemie} \textbf{410} 219.

\bibitem{mor04a} Morosan E, Bud'ko S L, Canfield P C, Torikachvili M S,
Lacerda A H 2004 \textit{J. Magn. Magn. Mat.} \textbf{227} 298.

\bibitem{mor05a} Morosan E, Bud'ko S L, Canfield P C 2005 \textit{Phys. Rev.
B} \textbf{72} 014425.

\bibitem{mor05b} Morosan E, Bud'ko S L, Canfield P C 2005 \textit{Phys. Rev.
B} \textbf{71} 014445.

\bibitem{kat04a} Katoh K, Mano Y, Nakano K, Terui G, Niide Y, Ochiai A 2004
\textit{J. Magn. Magn. Mat.} \textbf{268} 212.

\bibitem{tro00a} Trovarelli O, Geibel C, Cardoso R, Mederle S, Borth R,
Buschinger B, Grosche F M, Grin Y, Sparn G, Steglich F 2000 \textit{Phys.
Rev. B} \textbf{61} 9467.

\bibitem{kac00a} Kaczorowski D, Andraka B, Pietri R, Cichorek T, Zaremba V I
2000 \textit{Phys. Rev. B} \textbf{61} 15255.

\bibitem{bud04a} Bud'ko S L, Morosan E, Canfield P C 2004 \textit{Phys. Rev.
B} \textbf{69} 014415.

\bibitem{bud05a} Bud'ko S L, Morosan E, Canfield P C 2005 \textit{Phys. Rev.
B} \textbf{71} 054408.

\bibitem{mor05c} Morosan E, Bud'ko S L, Mozharivskyj Yu, Canfield P C 2005
arXiv:cond-mat/0506425.

\bibitem{and75a} Andersen O K 1975 \textit{Phys. Rev. B} \textbf{12} 3060.

\bibitem{and84a} Andersen O K, Jepsen O 1984 \textit{Phys. Rev. Lett.}
\textbf{53} 2571.

\bibitem{bar72a} von Barth A, Hedin L 1972 \textit{J. Phys. C} \textbf{5}
1629.

\bibitem{zim67a} Ziman J M 1967 \textit{Electrons and Phonons} (London:
Oxford University Press).

\bibitem{sch97a} Schnelle W, Poettgen R, Kremer R K, Gmelin E, Jepsen O 1997
\textit{J. Phys.: Cond. Matt.} \textbf{9} 1435.

\bibitem{and70a} Andersen O K 1970 \textit{Phys. Rev. B} \textbf{2} 883.

\bibitem{gri81a} Grimvall G 1981 \textit{The Electron-Phonon Interaction in
Metals} (Amsterdam: North-Holland Publishing Company).
\end{thebibliography}
\end{document}